\documentclass[%
superscriptaddress,
nofootinbib,
amsmath,amssymb,
aps,
twocolumn,
prb,
]{revtex4-2}
\usepackage{graphicx}
\usepackage{dcolumn}
\usepackage{bm}
\usepackage{xcolor}
\usepackage{tabularx}
\usepackage{amsmath}
\usepackage{gensymb}
\usepackage{upgreek}
\usepackage{siunitx}
\usepackage{array}
\usepackage{color,soul}

\usepackage{float}
\usepackage{enumitem}
\usepackage{gensymb}

\begin{document}
	
\title{Artificial spin ice systems on single edge length tilings}

 
\author{E. Weightman}
\affiliation{Department of Physics, University of Liverpool, Liverpool L69 3BX, United Kingdom}

\author{S. Coates}
\affiliation{Department of Physics, University of Liverpool, Liverpool L69 3BX, United Kingdom}


\begin{abstract}
Artificial spin ices (ASIs) are decorations of single domain nanomagnets on geometric lattices allowing for study of magnetic interactions on easily varied lattice geometries. We extend the study of ASIs on single rhomb tilings by simulating an ASI with the aperiodic hexagonal $H_{\frac{1}{2}\frac{1}{2}}$ tiling as the base structure, and the periodic dice lattice for additional comparison. Using a point dipole metropolis Monte Carlo approach, we cool our arrays of nanoislands to form a low energy configuration,  assessing individual vertex excitations as identified through complementary micromagnetic simulations. Using the energy and charge landscape of final spin states, we propose a potential ground state spin configuration for the aperiodic system.
\end{abstract}

\maketitle

\section{Introduction and motivation}

\begin{figure*}[t] %
	\label{fig:figure1}
	\begin{center}
		\includegraphics[width=\textwidth]{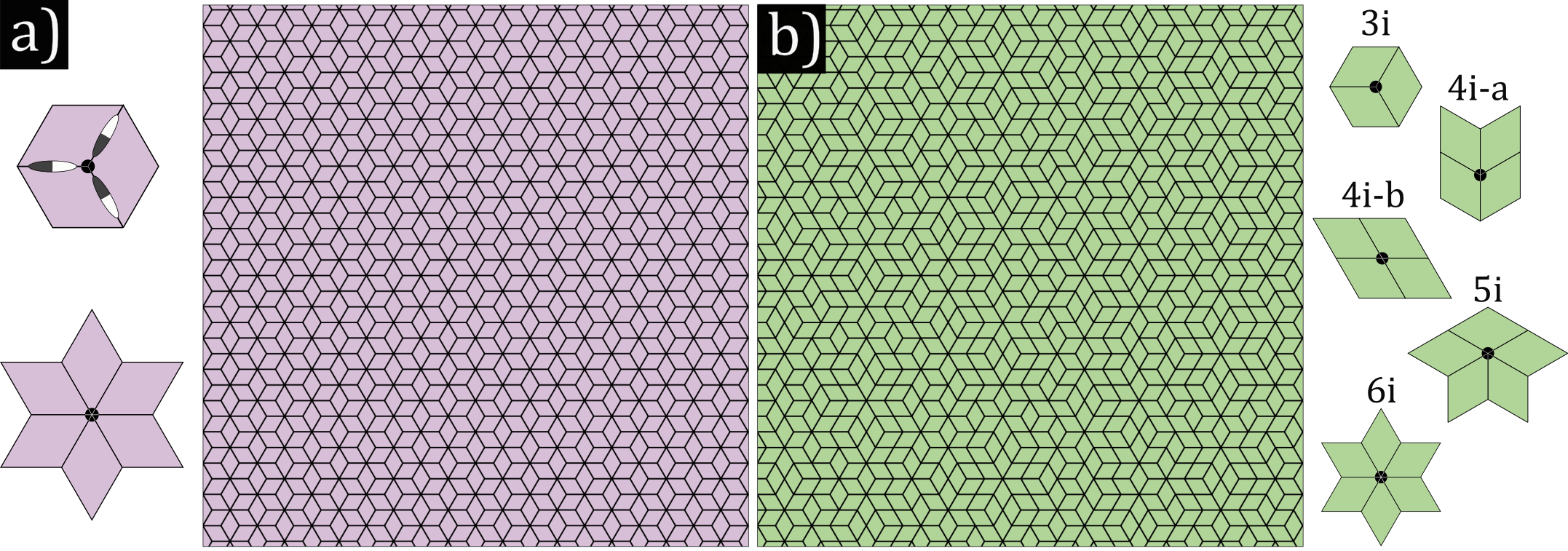} 
	\end{center}
	\caption{The two tilings used as artificial spin ice lattice geometries. The periodic dice tiling \textbf{(a)} with two vertex types of coordination 3 and 6, and the aperiodic $H_{\frac{1}{2}\frac{1}{2}}$ tiling \textbf{(b)}, a hexagonal golden mean tiling with 5 vertex types, coordination 3, 4, 5 and 6, which are labelled with respect to their island decoration (i). Both consist of a single diamond rhombus tile.} 
\end{figure*}

Artificial spin ices (ASIs) are novel structures consisting of arrays of needle-like, monodomain nanomagnets which decorate a pre-determined pattern \cite{P18wang2006artificial, P92gilbert2016emergent, P2shi2018frustration}. Initially explored as a model of physical spin ice materials, the geometry of an ASI array will typically govern a certain level of magnetic frustration in the system, broadly described as either vertex or topological. Vertex frustration occurs when nanomagnets meet and, despite being in their lowest energy configuration, will be `unhappy' as each of the magnets cannot align exactly antiparallel to each of its neighbours. An example is shown in Figure \ref{fig:figure1}(a) for a vertex with 3 islands, where each cannot be oriented opposite to both neighbours, thus causing frustration. On the other hand, topological frustration describes when it is not possible for all vertices to exist in their lowest energy state due to the global geometry of the system. Tuning such frustrations by both studying and altering the vertex arrangements in ASIs has led to the observation of many emergent phenomena, such as magnetic monopoles \cite{P93ladak2010direct}, reversal dynamics \cite{P5brajuskovic2016real}, and complex domain structures \cite{P91montaigne2014size}.

Two-dimensional tilings are often used as a framework to study a variety of lattice geometries, as they have well defined vertex structures and local environments. Studies of ASIs using periodic tilings as a base geometry have been extensive, with a particular focus on square \cite{P18wang2006artificial,P11drisko2017topological, P37brevis2021topological, P79porro2013exploring, P80zhang2013crystallites, P81ostman2018interaction, P82kapaklis2012melting} and Kagome \cite{P7li2022geometric, P9hugli2012artificial, P14canals2016fragmentation, P15rougemaille2011artificial, P84bhat2016magnetization, P85qi2008direct, P86moller2009magnetic, P87gartside2018realization} type tilings.

Conversely, the use of aperiodic structures, which possess no unit cell or translation symmetry, has been limited, with a few examples of Penrose tiling structures explored \cite{P2shi2018frustration, P3kwon2022searching, P5brajuskovic2016real, P6bhat2013controlled}. As both local and global properties of ASIs are dependent on allowed spin arrangements at vertices and their distribution, aperiodic tilings represent a rich playground with which to explore ASI systems; aside from their novel structure, their vertex properties are often highly tunable. As an example, manipulating the shifts applied to the grids in dual-grid based tilings can either subtly or dramatically change the overall structure (and hence vertex distribution) of a given tiling `family' \cite{P13coates2024hexagonal}. We are therefore motivated two-fold: first, to broaden the scope of explored aperiodic ASIs. Second, to demonstrate how altering vertex types and their distributions in an aperiodic arrangement influences magnetic properties, in comparison with a well-studied periodic system.
Here, we explore theoretical ASIs on two systems -- the periodic `dice' lattice and the single length $H_{\frac{1}{2}\frac{1}{2}}$ tiling -- which share the same building block: a single edge length diamond rhombus. Figure \ref{fig:figure1}a shows a patch of the periodic dice tiling, alongside its two vertex types of coordination 3 and 6 respectively. Previous experimental studies of this lattice have revealed charge screening behaviour as the system equilibrates \cite{P76farhan2016thermodynamics} -- a direct consequence of these vertex types. Figure \ref{fig:figure1}b shows a patch of the single length $H_{\frac{1}{2}\frac{1}{2}}$ tiling  \cite{P13coates2024hexagonal,P78Coates2025designing}, where here each of the unique tiles of the original tiling is coloured identically. The tiling contains 5 vertex types: it shares the 3 and 6 types of the dice lattice while introducing two distinct vertex types of coordination 4, and a 5 type. 
We investigate these specific systems for a number of reasons: first, they enable a direct comparison of equivalent geometric features, discussed below, while minimizing the spatial discrepancies that typically arise when contrasting periodic and aperiodic geometries (e.g. multiple edge lengths, distance between opposing vertices etc.). Similarly, it has recently been shown that these two systems can be directly related by simply rotating specific clusters of tiles \cite{P78Coates2025designing}. Therefore, our work represents an initial case study on a whole family of related rhombic systems, with scope for further work, e.g. inducing disorder by changing the percentage of rotated clusters etc. Lastly, we are motivated by the work of \cite{P72cote2023direct}, where a series of ASIs with varying degrees of structural order -- constructed using the same diamond rhomb -- were investigated experimentally. In this work, we use Monte Carlo and finite-difference simulations to explore the theoretical ground state of the $H_{\frac{1}{2}\frac{1}{2}}$ ASI, compare its magnetic properties to the dice ASI, and introduce a potential analytical ground-state solution for the aperiodic system.

\section{Methods}
 
 \subsection{Tilings}
Both tiling structures were generated using a generalisation of the dual grid method \cite{deBruijn81, deBruijn86, Socolar85, Gahler86, Ho86, Rabson88, Rabson89, Lifshitz05,Luck1993basic,Ingalls1993octagonal}, which we briefly explain here. We start with an infinite set of equally spaced parallel lines, a grid. The lines of this grid are perpendicular to some grid vector $\textbf{g}^{(j)}$, and are separated by the magnitude of this vector. Introducing additional grids at some non-zero orientation (via new grid vectors) creates intersection points between the two sets of overlapping grid lines. Associated with the set of grid vectors $\textbf{g}$ are the tiling vectors, $\textbf{t}$, which are identical in orientation, but not necessarily magnitude. Each intersection point then corresponds to a tile in the resultant tiling, whose edges are the tiling vectors associated with the families of the intersecting grid lines. The intersection points for a given set of $\textbf{g}$ and $\textbf{t}$, and hence the tiling, can then be manipulated by shifting each grid along the direction of its vector, by some fraction of its vector magnitude.

The tile and grid vectors for both tilings we explore can be described in terms of the unit vectors:
\begin{equation}
	\textbf{a}^{(j)} = (\cos{\frac{2\pi(j-1)}{3}}, \sin{\frac{2\pi(j-1)}{3}})
\end{equation}

where $j = 1, 2, 3,...,6$. For the dice lattice, the tile and grid vectors are simply \textbf{a}. For the single edge $H_{\frac{1}{2}\frac{1}{2}}$ tiling we apply a scaling for the grid vectors only, where the magnitude of vectors $j = 1, 2, 3$ is set to the golden mean, $\tau$. The overall effect is to aperiodically distribute the same diamond rhombus as used in the dice lattice, as discussed in \cite{P78Coates2025designing}. Briefly, changing the scaling factor between the two families of grid vectors alters which grids intersect and where; a `short' grid vector will induce more intersection points as the grid lines are more densely distributed, and vice versa. An irrational scaling factor then ensures an aperiodic distribution of intersection points. To obtain the dice-like tiling structures for both tilings, we separate the grid vectors into two families ($j$ = 1--3 and $j$ = 4--6), and sum the shifts of the grids in each family to 0.5, such that we create tilings with hexagonal symmetry.

\subsection{Monte Carlo simulations}
The two lattice geometries were decorated with `islands' placed on the midpoints of rhomb edges. Here, the islands are considered to be monodomain nanomagnets, pointing toward and away from the two tile vertices which they lie between. Interactions were calculated between islands using a Metropolis Monte Carlo algorithm \cite{P83metropolis1949monte}, with the following Hamiltonian:

\begin{equation}
	H = -J  \; \sum_{\left<ij\right>} \: S_i \cdot S_j + \frac{D}{2} \: \sum_{ij} \left[ \frac{ S_i \cdot S_j}{r_{ij}^3} - \frac{3 \cdot (S_i \cdot \bar r_{ij}) \cdot (S_j \cdot \bar r_{ij})}{r_{ij}^5} \right]
\end{equation}

In this way, the islands are modelled as point dipoles, except in nearest neighbour cases where an additional term is added to account for an enhanced coupling at smaller distances.  Studies such as \cite{P14canals2016fragmentation, P15rougemaille2011artificial} have found this reliable for reproducing experimental phenomena. The interaction is governed by two coupling constants, $J$ and $D$, where $J$ is the constant governing the additional term applied only in nearest neighbour magnetic interactions, chosen to be negative for anti-ferromagnetic coupling, and $D$ is the constant governing dipolar interaction, which is applied to all spin interactions. We classify nearest neighbours as points which lie within one edge length of each other, and truncate the range of dipolar interactions to include only those at the nearest 5 unique distances.

We calculate realistic values for the coupling constants for a specific physical system commonly used experimentally: permalloy Ni$_{80}$Fe$_{20}$ nanoislands which have a saturation magnetisation of $8.6\times10^5$Am$^{-1}$ \cite{P5brajuskovic2016real}, dimensions 400nm$\times$100nm$\times$10nm. The islands decorate the edges of the tiles but do not meet at the vertices, therefore, there is a `gap' of 50nm from the tip of the island to the vertices of the tile. Considering this, we use $\left| J/D \right|$ as: 1.95. 

We cool the spin system through a a logarithmic series of 180 temperatures, undertaking  $10^6$ Monte Carlo steps in each and sampling with a Metropolis Monte Carlo algorithm. In each step, a random spin flip is proposed. The new spin may be accepted firstly, if energetically favourable, $\Delta E < 0$, or secondly, accepted with the probability $P=e^{-\frac{\Delta E}{T}}$. The $k_{B}$ term is omitted here as our system is unitary. As the temperature is reduced, this induces a low energy spin configuration in the system. 
We use temperatures scaled by the nearest neighbour coupling constant, J, such that the highest energy interactions are normalised to 1. We simulated an ensemble of 15 systems for both the periodic and aperiodic cases, each containing approximately 3114 islands in the ASI, with each system initialised with a different randomised spin state. The total energy of the system was sampled at increments of 10\% of the total number of attempted spin flips (100 samples per temperature), and averaged at each temperature. We simulated both ASIs with open boundary conditions to more accurately represent an experimental physical array, where we found edge effects to have a negligible contribution to the energetics or ordering of the systems. As such, we do not include them in our discussion of these ASIs.

\subsection{Vertex energies}

To supplement the Monte Carlo simulations of the arrays, we conduct a vertex level analysis of each system's energy states. Using the MuMax3 software package \cite{P61vansteenkiste2014design}, we calculate the energy of all spin configurations of each of the five vertex types in a physical spin system, allowing us to identify the energy states of each vertex in the Monte Carlo simulation results. In this method, each magnetic island is fractionalised into discrete micromagnetic elements in a mesh which are individually resolved with a finite difference method. As such, we assess the energetic properties at the vertex sites, which is in turn driven by their coordination and respective populations across the array. We model islands of the same properties as discussed previously, permalloy nanoislands of dimensions 400nm$\times$100nm with a constant magnetisation imposed in the z-axis to model a 2D system. From this, we are able to identify the comparable energy states of each vertex configuration and use this to identify the energy states of vertices in the low temperature state of the simulations.

\section{Results and discussion}

In this section, we first discuss systemic properties sampled across the Monte Carlo simulations before using vertex energy calculations to statistically characterise the average final state for each ASI. Then, we show how resultant finite magnetic charges may lead to a potential analytical ground state solution for the aperiodic ASI, before discussing magnetic monopoles at domain boundaries. We note for further discussion that, in the literature, a monopole is defined as a site of \textit{maximum} finite charge at a coordination site, not necessarily any given charge.

To aid comparison to previous studies of rhombic ASIs, we apply an order parameter notation for each tiling, previously defined for rhomb tilings of mirrored and rotated tile arrangements \cite{P72cote2023direct, P71stannard2012broken}. While it has been shown that this order parameter does not necessarily indicate the level of magnetic ordering, it is convenient for characterisation and comparison. The order parameter is given as: 

\begin{equation}
 \psi = \frac{0.608 n_{rot} - 0.392 n_{mirror}}{0.608 n_{rot} + 0.392 n_{mirror}}. 
\end{equation}

\noindent which, for our periodic and aperiodic structures are -1 and -0.36 respectively, such that they are subsequently denoted as $\psi_{-1}$ and $\psi_{-0.36}$.

\subsection{System behaviour}

\begin{figure*} %
	\begin{center}
		\includegraphics[width=0.95\textwidth]{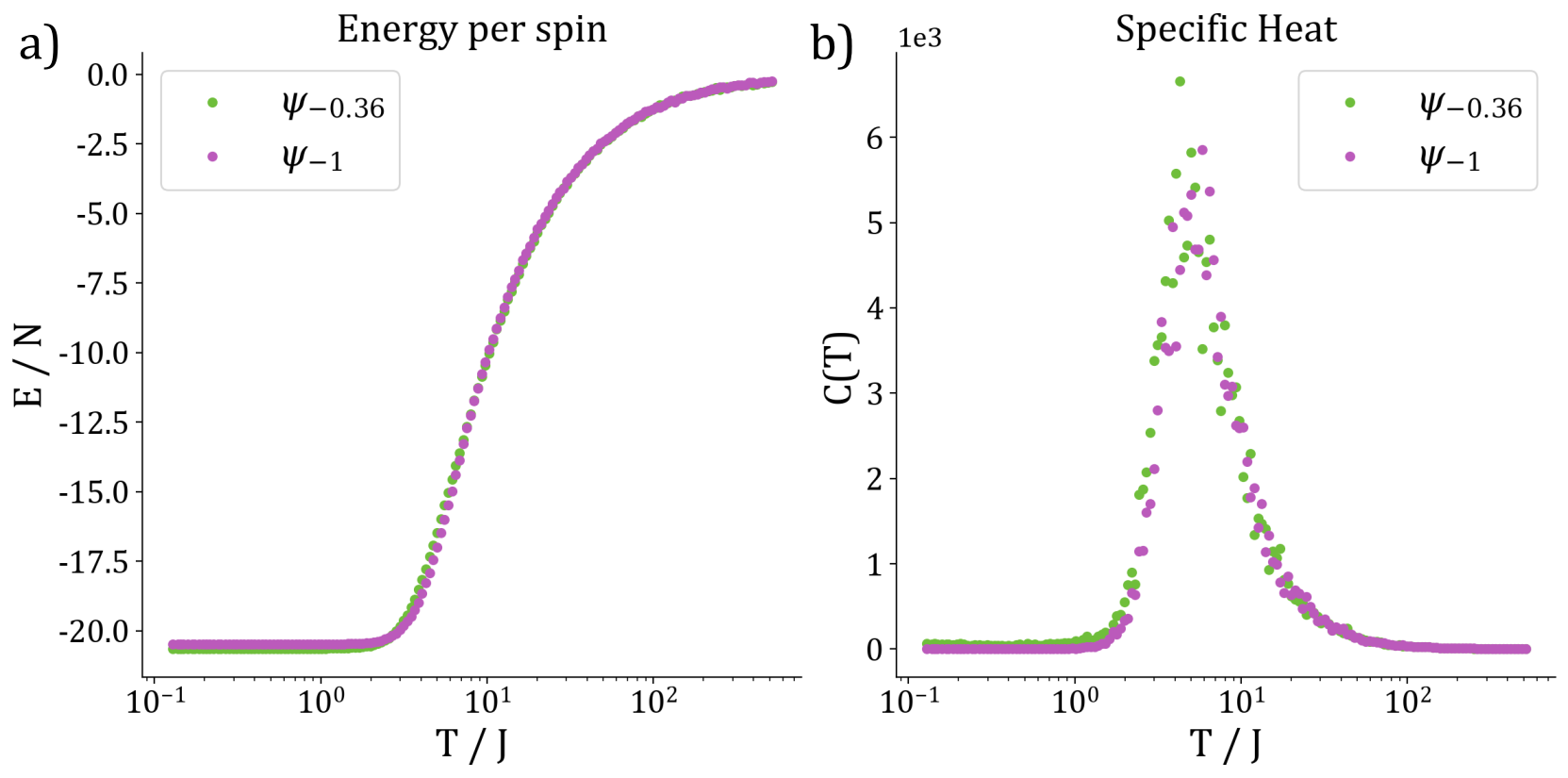} 
		\end{center}
	\caption{\textbf{(a,b)} Total system energy per spin and specific heat, respectively, averaged over 15 simulations of 3106 and 3114 nanoislands in the $\psi_{-1}$ and $\psi_{-0.36}$ geometry.}\label{fig:EMCX}
\end{figure*}

Figure \ref{fig:EMCX}(a) shows the energy per spin of the $\psi_{-1}$ and $\psi_{-0.36}$ spin systems averaged over the ensemble. As the temperature is reduced, both systems freeze into a lower energy state at comparatively similar temperatures, which is also evident in the calculated average acceptance rate, shown in Appendix \ref{app:acceptance rates}. As expected for a rather simplistic cooling regime, the rate is maximised at higher temperatures then reduces to approximately 50\% at the critical temperature, before reducing to 0\% acceptance as the system freezes and no more spin flips are accepted. In both cases then, we do not find the true absolute ground state; the single-spin-flip dynamics become kinetically trapped once the acceptance rate falls to zero. This is discussed further on with analysis of the excitations present, and we note that more sophisticated and acceptance-rate-adaptive cooling will be explored in future work.

However, we can quantify the accuracy of our simulations by calculating the residual entropy of the systems and comparing to known values. To do so, we have calculated the specific heat from our energy sampling as: 
\begin{equation}
	C(T) = (\langle E^2 \rangle - \langle E \rangle ^2)/T^{2}
\end{equation}
which is shown for both systems in Figure \ref{fig:EMCX}(b). Then, the residual entropy for each final spin state is:
	\begin{equation}
		S_r = S_{max} - \int_{T1}^{T2}\frac{C(T)}{T}dT
	\end{equation}
	
where S$_{max}$ is the maximum entropy at high temperature; for a system of $N$ spins with 2 possible orientations per spin, this is simply $N\ln{2}$, or $\ln{2}$ for entropy per spin (where we take $k_b = 1$).
	
For the $\psi_{-1}$ geometry, $S_r$ = 0.205. Normalising this by the frequency of the 6i vertices gives an approximation of the entropy contribution from the 3i vertices only, such that $S_r$ = 0.311. This allows us to compare directly to the triangular Ising antiferromagnet, where $S_r$ = 0.323 \cite{wannier1950antiferromagnetism}. The close agreement between these values (within 4$\%$) suggests that our simulated annealing protocol approaches the expected ground-state degeneracy of the 3i sublattice with reasonable fidelity. We note that our normalisation assumes the 3i and 6i sites contribute independently to the total entropy -- correlations between adjacent sites are not accounted for, and may introduce a small systematic offset. We find that for $\psi_{-0.36}$, $S_{r}$=0.247, higher than $\psi_{-1}$ and lower than the triangular Ising anti-ferromagnet, indicating a system which is entropically more degenerate than the dice, but less than the triangular.

\begin{figure*} %
	\begin{center}
		\includegraphics[width=0.95\textwidth]{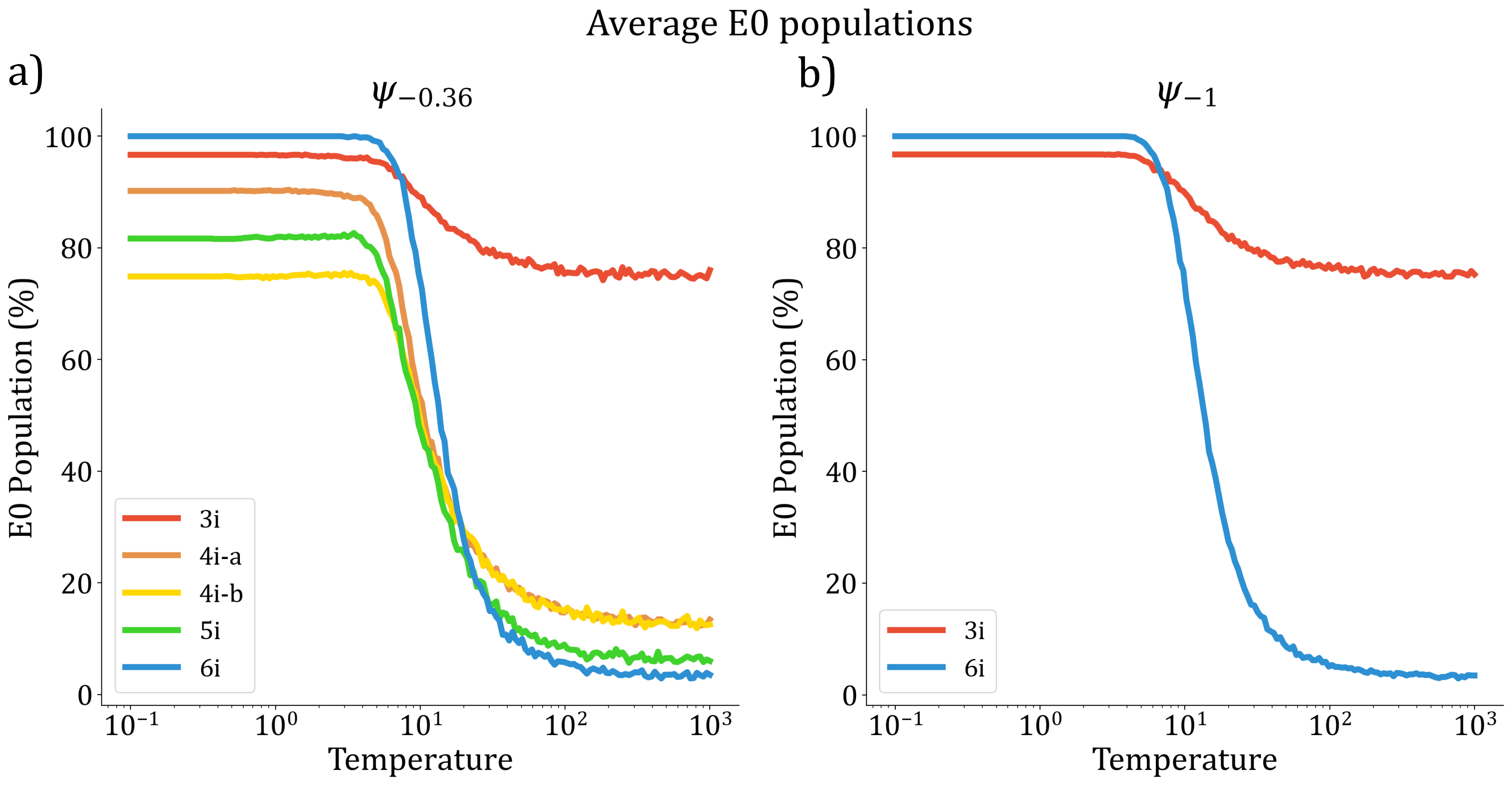} 
	\end{center}
	\caption{E0 populations for each vertex type in the $\psi_{-1}$ and $\psi_{-0.36}$ geometries as the temperature of the system is reduced.}\label{fig:E0pop}
\end{figure*}

To demonstrate how each system finds global low energy states, Figure \ref{fig:E0pop} shows the average local energy state population for each vertex type as the systems are cooled. The initial populations are probabilistic, determined purely by the proportion of all configurations which are E0. For example, the 3i vertex type has 2$^3$ possible configurations, of which 6 are E0, thus the initial E0 population is approximately 75\%. Similarly, the 6i vertex type has 2$^6$ possible configurations of which 2 are E0. As such, the initial population is approximately 3\%.

In both geometries all E0 populations begin in their probabilistic frequency, increasing as the temperature is decreased, before reaching a plateau at which the spins have frozen. Given that the E0 populations do not reach 100\% for all vertex types, the low energy states of the spin systems must consistently possess excitations. The 6i vertices in both geomtries reach 100\% E0, and plateau first compared to other vertex types. The percentage of E0 3i is also high, $\sim$ 97\% in both geomtries. In the $\psi_{-0.36}$ system, the additional vertex types plateau at similar temperatures reaching different E0 population levels. The 4i-a vertices have the highest E0 population of 90\%, with 5i having 82\% and 4i-b being the least energetically happy with an E0 population of 75\%.

\begin{figure*}[t] %
	
	\begin{center}
		\includegraphics[width=0.98\textwidth]{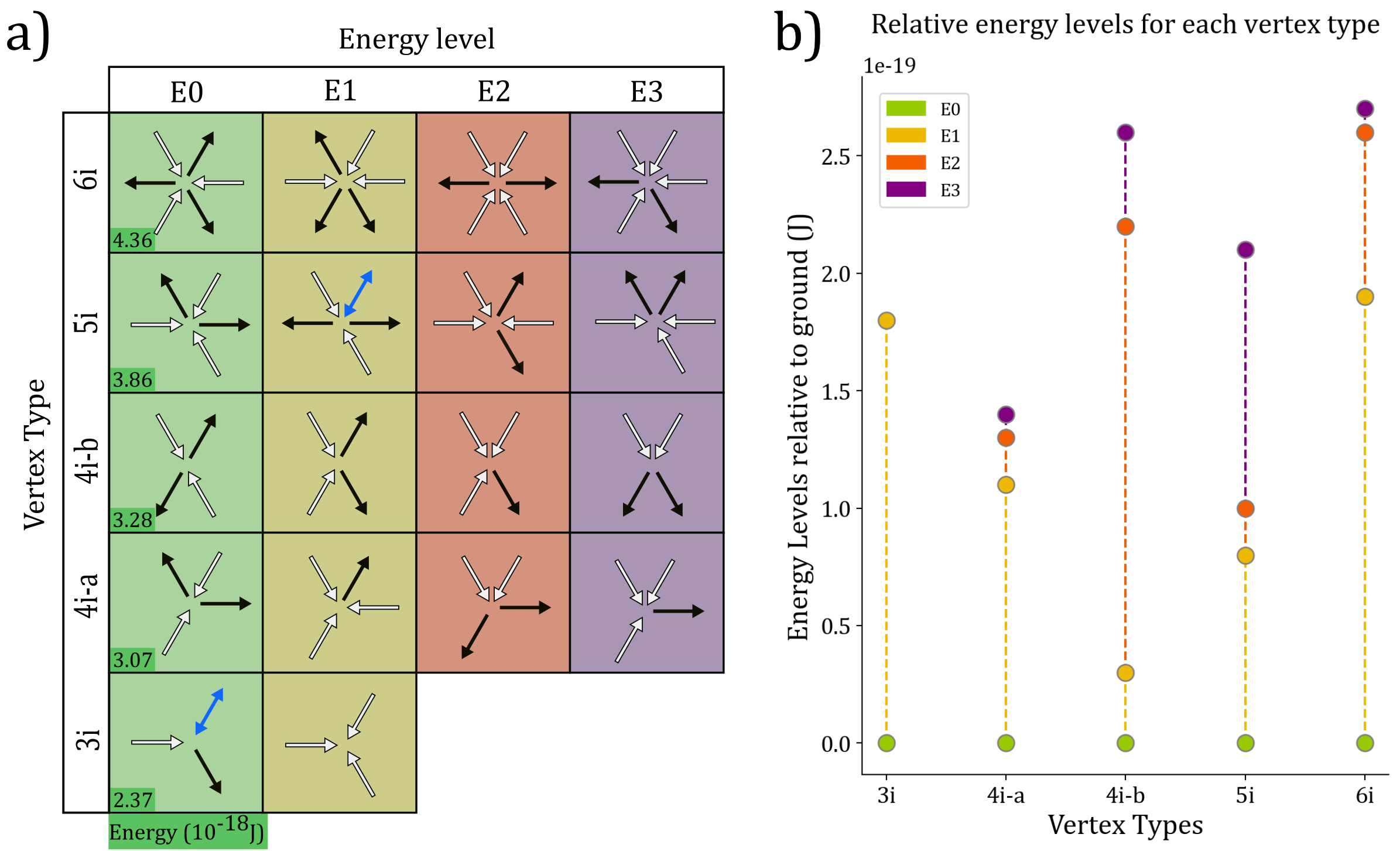} 
	\end{center}
	\caption{\textbf{(a)} The four lowest energy spin configurations for each vertex type, calculated in micromagnetic simulations. E0 denotes the lowest energy configurations with spin directions obeying the ice rules. Each configuration is at least two-fold degenerate, and some states have many equivalent analogues, such as 4i-b E2, where the out spin can occupy any of the four islands shown. Blue arrows represent spins which may degenerately point in either direction. The configurations are thus four-fold degenerate \textbf{(b)} The first 3 excited energy states for each vertex type relative to the ground energy level.}\label{fig:VEN}
\end{figure*}
\subsection{Vertex energy states}

Figure \ref{fig:VEN}(a) shows the lowest 4 energy configurations for each vertex type obtained from micromagnetic simulations, labelled from E0 to E3, where we have omitted the higher energy levels for clarity. White or black arrows indicate the direction of the island spin, in and out respectively. Blue arrows represent spins which may degenerately point in either direction. Each state, except those with blue arrows, is two-fold degenerate such that all spins can be flipped without changing their energy; for conciseness we only show one of many equivalent states at each level -- such as 4i-b E2, where the out spin can occupy any of the four islands shown. Those with blue arrows possess four-fold degeneracy in that all spins may flip, and this spin may point in either direction in both states. The E0 configuration for each vertex type obeys the ice rules \cite{P77bernal1933theory} i.e. in-out-in-out for consecutive spins. Higher energy configurations attempt to prioritise anti-parallel spins where islands are closely separated -- for example, in the E1 state of the 4i-a vertex, one island sits further from the other three. This leads to the three islands coupling strongly in an anti-parallel fashion, with the `lone' island breaking the ice rules. The absolute energy differences of the vertex excitations, measured relative to their lowest energies (set to zero), is shown in Figure \ref{fig:VEN}(b). We note for future discussion that the 4i-b vertex requires the least energy to be excited into its E1 state.

\begin{figure*}[t] %
	
	\begin{center}
		\includegraphics[width=\textwidth]{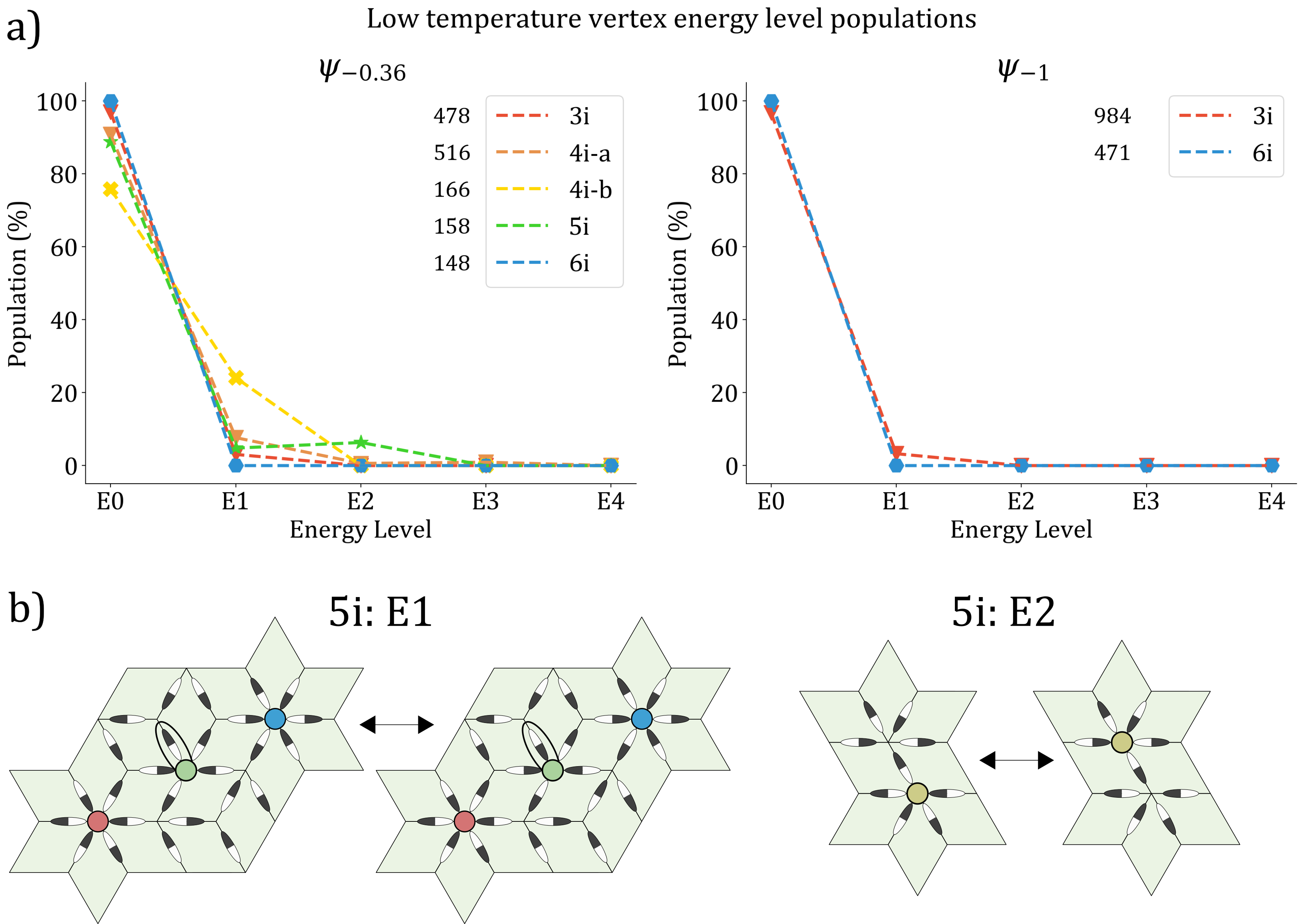} 
	\end{center}
	\caption{\textbf{(a)} Low temperature energy level populations for each vertex type for $\psi_{-1}$ and $\psi_{-0.36}$, with the vertex populations shown alongside the legend. \textbf{(b)} Schematic diagram showing how the excited states of the 5i vertices arise. Left: the 5i E1 state, where a green circle highlights the excited vertex, and red/blue circles indicate two anti-parallel magnetic domains propagated by the 6i vertices. An oval shows how the middle spin can flip without altering the energetics of the local environment. Right: the 5i E2 state, in which the excited state `hops' between two connected 5i vertices after each flip.}\label{fig:enpop}
\end{figure*}

Using the calculated energy states, we have tracked the frequency of the populated energy levels in the final simulated state across both the $\psi_{-1}$ and $\psi_{-0.36}$ systems, shown in Figure \ref{fig:enpop}(a). Generally, both systems follow basic thermodynamic reasoning: higher-energy states are less likely to be populated. For $\psi_{-1}$, all of the 6i vertices reached their lowest energy configuration, while 97\% of 3i vertices were found in E0, such that the remaining 3\% were fully excited, forming a monopole of charge $\pm$3. These statistics are matched for the 3i and 6i vertices of $\psi_{-0.36}$. However, we note two highlights from Figure \ref{fig:enpop}(a). First, it shows that the 4i-b vertices represent the lowest proportion of vertex types in their E0 state (75\%), with a correspondingly high share of excitations to the E1 state. As mentioned earlier, this is the lowest energy excitation across all vertex types. This suggests that the route to a low energy ground state for $\psi_{-0.36}$ is mediated via these excitations, as opposed to a lower frequency of higher energy states. Second, the 5i vertices are the most prevalent of the high-energy states, occupying their E1 and E2 levels 4.4\% and 5.8\% respectively. These relatively high occupations can be explained by examining common local environments of 5i vertices. 

Figure \ref{fig:enpop}(b) shows two such environments, where on the left side we see an E1 state, and on the right side an E2 state. Relevant tile edges are decorated with islands, as in Figure \ref{fig:figure1}(a). For the E1, the 5i vertex is indicated by a green circle -- we note that in Figure \ref{fig:VEN}(a), its middle spin is degenerate. Nearby, there are two 6i satisfied vertices whose spins are anti-parallel with respect to one another, representing two different magnetic domains, which we indicate by red and blue circles. These domains are coupled via the surrounding lowest energy states of the two 4i-a and 3i islands, which leaves the 5i vertex `pinned' in its E1 state. Flipping either all of the spins on the red or blue site would cost too much energetically -- both locally and on the surrounding vertices. The result is that the middle 5i spin, highlighted by an oval and connected to a 3i vertex, can flip at will with no change to the energy of the system -- as the 3i is also satisfied by either spin direction. Similarly, we find that the E2 state occurs when two 5i vertices which belong to two different magnetic domains are connected. In this instance, one domain would consist of out-in-out-in-out spins, and the other vice versa. The excited state is indicated by a circle, where flipping the connecting spin de-excites one vertex to its ground state, and excites the other to E2. We can tie these observations to our previous discussion of the parameters in Figure \ref{fig:EMCX}: both of these environments are entropically stable, yet represent a change in the magnetisation of the system as the spins flip between each state. Studying the dynamics of these flips and their frequency across the tiling will be a topic of future work.

\subsection{Magnetic States}

\subsubsection{Finite magnetic charges as a route to ground state solution}

The ground state solution of the $\psi_{-1}$ system is known and simple to verify -- both 6i and 3i vertices can simultaneously occupy their E0 levels in such a way that consistently decorates the dice lattice (or even with multiple domains) in a well defined low energy state of two fold degeneracy. On the other hand, the solution for $\psi_{-0.36}$ is not immediately obvious given the enhanced complexity in the structure. To find an idealised ground state, it is instructive to analyse the average final spin state structure. However, averaging the spin directions of each island introduces incoherence: degeneracy can cancel the average spin at a site. Fixing a given spin relative to the array resolves this issue locally, but the presence of multiple domains in each simulation state then introduces noise. 

\begin{figure*} %
	\begin{center}
		\includegraphics[width=\textwidth]{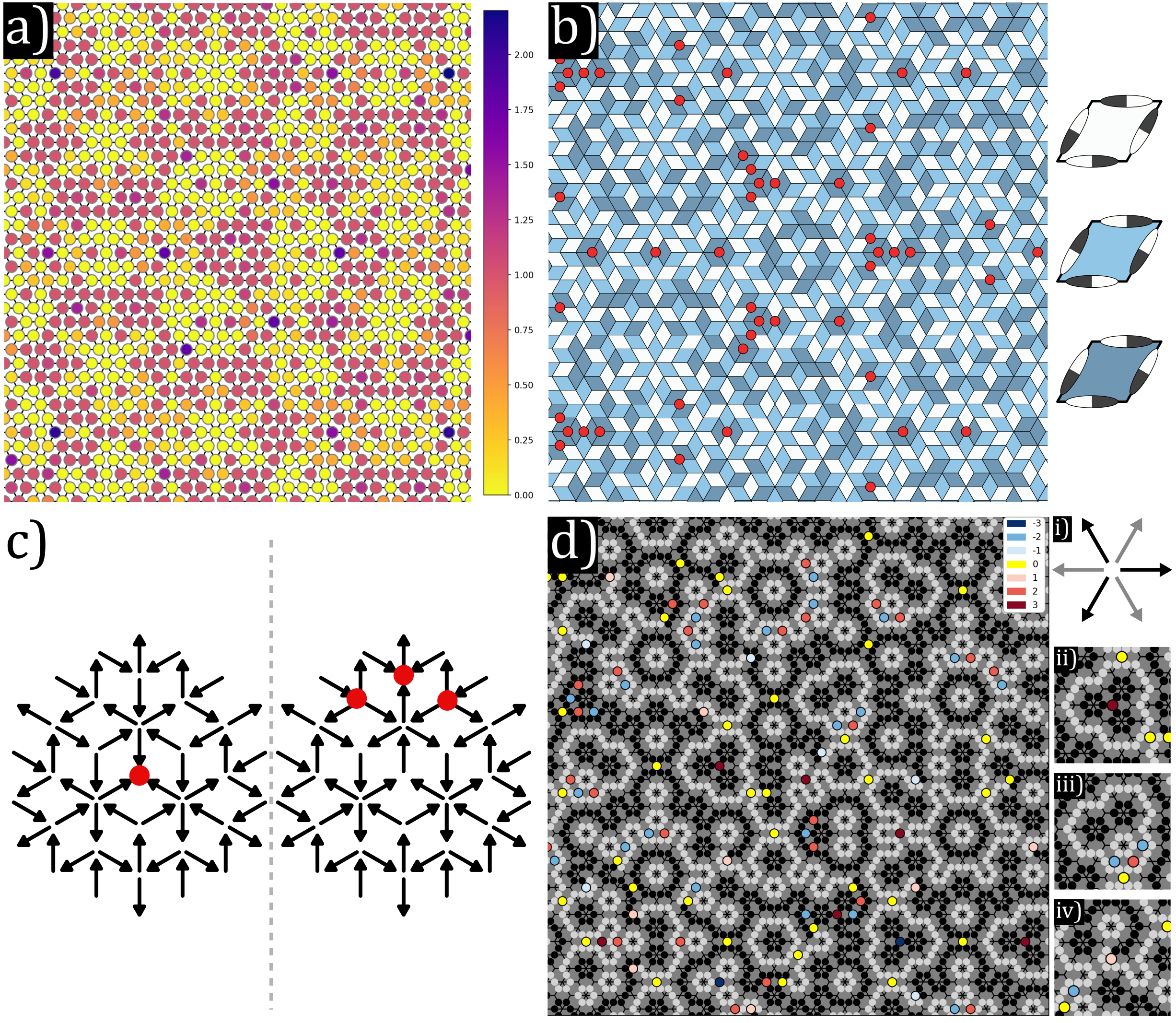} 
	\end{center}
	\caption[width=0.9\textwidth]{\textbf{(a)} Average finite magnetic charge on each vertex in $\psi_{-0.36}$, overlaid on the tiling geometry. The colours of the circles represent the average of the sum of charges pointing toward and away each vertex with darker circles indicating where greater finite charge is found. \textbf{(b)} The H$_{\frac{1}{2}\frac{1}{2}}$ tiling, built using three decorated tiles (right), used to build the ground state. Red circles indicate excited vertex states. \textbf{(c)} Simulated systems show a lower percentage of 3i excitations compared to the analytical solution, which can be explained by the `trapping' of low energy states. On the left-hand side, the 3i monopole is pinned by three aligned 6i vertices. On the right-hand side, the top-most 6i is flipped, which de-excites the 3i, but excites the three 4i-a vertices. At relatively high T, the 6i vertices freeze, which traps the 3i into a local minimum state, but increases global energy.  \textbf{(d)} Circles decorate the mid-edges of the $\psi_{-0.36}$ geometry, corresponding to the spins in the system. Circles are coloured according to the two colour domains: black indicates spins of angles 0\degree, 120\degree and 240\degree, and lighter grey spins 60\degree, 180\degree and 300\degree: as shown in \textbf{i}. Excited states and monopoles are also shown at vertex sites, with their charge indicated by the colour-scale legend. Inset panel \textbf{ii} shows a monopole arising from the ground state solution, while \textbf{iii,iv} highlight excitations caused by colliding domain behaviour.} \label{fig:4parter}
\end{figure*}

Instead, we assess the absolute finite magnetisation of the average final state, wherein each island is considered as a charge of 1, with +1 pointing toward a vertex, and -1 oriented away. We then sum the charges at each vertex, and find the absolute value to ensure that charged states are preserved in the average, rather than being cancelled by spin degeneracy, as individual finite charges (independent of polarity) may exist at consistent sites given the global geometry we enforce. Given this `charge map', we can then attempt to decorate $\psi_{-0.36}$ with its lowest energy states in such a way that satisfies this average final magnetisation state. 

Figure \ref{fig:4parter}(a) shows the visualisation of these absolute finite magnetic charges for $\psi_{-0.36}$, averaged across 15 simulations. Each circle represents a vertex in the array, and is coloured according to its average charge, from 0 to 2.2. We find that regions of zero or near-zero charge correspond to even-coordination vertices, higher values arise from odd-coordination vertices and excited 4i-a types, and the largest charge values occur when 3i vertices forms a monopole. Statistically, we also find this to be true: by cross-referencing the percentages energy level occupancies in Figure \ref{fig:enpop}(a) with the calculated states in Figure \ref{fig:VEN}, we find globally that all 6i vertices are charge 0, all 5i vertices have charge +1, and all 4i-b vertices are charge 0, regardless of excitations. Therefore, there are variations in charge for the 3i and 4i-a vertices only. 

Using this magnetic state representation, we are able to derive a provisional geometric ground state model, utilising the clear symmetry of the finite charge distribution and the regions of zero or near-zero charge. First, we fix the central 6i vertex to be in one of its two degenerate ground states. Then, guided by the local charge distribution, we attempt to propagate the lowest possible energy configurations of connected vertices by simply overlaying these states onto the tiling. Quickly, we find that it is possible to continue this method by simply overlaying an additional set of three tiles with islands decorating their edges. Figure \ref{fig:4parter}(b) shows a patch of the resultant tiling, where the three tiles and their island decorations are highlighted adjacent. Red circles indicate excited states in the system, which solely occupy 3i and 4i-b vertices. Here, the 3i sites represent +/-3 monopoles, which the 4i-b vertices retain zero charge, but are excited into their E1 state. Attempting to remove these excitations is energetically costly given the local environment of each: the excited 3i vertices are surrounded wholly by either 6i vertices, or 4i-a vertices. To de-excite the 3i vertex would require excitation of more than one 6i or 4i-a which, as seen in Figure \ref{fig:VEN}(b), is much more energetically costly. Similarly, the excited 4i-b vertices are bounded by 4i-a vertices, which are again more costly to excite. As such, these in built excitations represent the smallest energetic cost given their local environment. Considering the decoration of these tiles, we can now re-introduce the polarity of charges into the system to find a provisional ground state. Picking one degenerate state arbitrarily, Figure \ref{fig:chargemap} shows the final polarised finite charge map. Here, white circles represent even vertices with a no charge, while yellow are the excited 4i-b vertices which also retain zero charge. Light red and blue circles have a value of +1 and -1 respectively, and the darker red and blue circles indicate a vertex of highest finite magnetic charge of +3 and -3 respectively. Given that the lowest energy state contains excitations, the $\psi_{-0.36}$ is, by definition, a topologically frustrated system. 

\subsubsection{Quantitative analysis of the ground state solution}

To quantitatively validate our proposed solution, we assess the total energy of both the average final simulation state and our analytical solution, in order to demonstrate that the latter yields a lower energy configuration. To do so, we first inspect and discuss the percentage of excited states across both systems. 
	
As previously shown in Figure \ref{fig:enpop}(a), no 6i vertices are excited in either system, as is to be expected given their coordination number and increased coupling. The excited populations of 4i-a, and 5i vertices are reduced from $\sim$8.3\% and $\sim$11.1\% respectively in the simulated system, to zero in the analytical, while the 4i-b are reduced from $\sim$24.1\% to 19.3\%. We note that in the simulations these states often appear at the boundary between equivalent yet opposing magnetic domains, discussed further below, which is necessarily accounted for in our solution. Perhaps surprisingly, the 3i vertices are excited slightly more often in the analytical solution (4.2\%) than in simulation ($\sim$3.1\%). However, as previously mentioned, some of the 3i monopoles are surrounded by 6i vertices. By inspecting the local environment of this particular arrangement, we see that a 3i can be `trapped' into its lowest state.
	
Figure \ref{fig:4parter}(c) shows two different scenarios for this environment, which consists of a central 3i, three connecting 6i, and twelve surrounding 4i-a vertices. On the left-hand side, the 6i vertices are aligned such that equivalent spins point in the same direction, pinning the 3i into a monopole. On the right hand side, the topmost 6i is flipped, which de-excites the 3i, but excites three 4i-a vertices at roughly double the energy cost. As Figure \ref{fig:E0pop}(a) shows, the 6i vertices are the first to freeze into their configurations at higher T. Locally, then, this right-hand motif is preferred at higher T, which freezes the 3i in its E0 state, at the cost of increasing the global energy.

Finally, then, we use the average vertex energy populations and the energies calculated from our micromagnetic simulations to calculate the total energies for the simulated and analytical systems for comparison. We find $4.527\times10^{-15}$J for the simulated and $4.521\times10^{-15}$J for the analytical systems respectively. The difference in the two energies is small, 5.77$\times10^{-18}$J, yet we can express it directly in terms of the additional excitations which occur between colliding domains. On average, there are 64 additional excitations in the simulated states, which gives $\sim0.9\times10^{-19}$J per excitation. According to Figure \ref{fig:VEN}(b), this corresponds roughly to one 4i-a E1 excitation, an E1 or E2 5i excitation, or multiple 4i-b excitations, all of which we routinely observe at the domain walls. We note however that the ground state solution we show has fundamentally been derived only over a small patch of the $\psi_{-0.36}$ system. Future work aims to verify this configuration as a true ground state by analytically solving the proposed charge distributions using substitution rules, with additional analysis utilising perpendicular space.

\subsubsection{Competing magnetic domains}

Using the provisional ground state solution we can better analyse domain behaviour observed in the final spin states -- relevant to, for example, the excited E1 and E2 states of the 5i vertices we observed. Figure \ref{fig:4parter}(d) shows an arbitrarily selected final spin state, where we represent each spin as a circle on the mid-edge of a tile. The spin circles are grouped into two distinct `colour families' based on the orientation of their spin, as indicated by the vectors in Figure \ref{fig:4parter}(d)i). The black circles correspond to the spins at the mid-edges of tiles, with angles 0\degree, 120\degree and 240\degree, and the lighter grey circles to angles 60\degree, 180\degree and 300\degree. We note that a ground state system of spins should not and could not exhibit a singular colour, trivially, as the geometry does not allow it. For example, Figure \ref{fig:4parter}(d)ii) shows a patch of $\psi_{-0.36}$ which explicitly belongs to one of the two ground states, where a +3 monopole 3i vertex (dark red) is surrounded by three black hexagons, which in turn is enclosed within a closed `loop' of grey circles. By flipping all of the spins (and hence their colour), we obtain the other ground state. Appendix \ref{app:groundcolour} shows a full colour map which represents one of the two degenerate ground states, for reference. 

Therefore, Figure \ref{fig:4parter}(d) serves to highlight the behaviour which arises when the two degenerate ground state solutions collide. The origin of each the two domains is simple: the strong coupling of satisfied 6i vertices means they are `locked in' before other vertex types as seen in Figure \ref{fig:E0pop}, as too much energy is required to excite and fully flip them. However, there is not a strong enough coupling between adjacent 6i vertices to ensure that they settle in the same spin configuration. Adjacent vertices then find their minima according to the orientation of these 6i sites, propagating until two anti-parallel ground states collide, resulting in excited states at the domain boundaries. These sites are indicated across Figure \ref{fig:4parter}(d), and are coloured from a scale of blue to red, representing positive to negative charge. Here, we find excited states and monopoles at two types of sites: those which exist on the proposed ground state(s), and those induced by the colliding magnetic domains. An enlarged example is shown in Figure \ref{fig:4parter}(d)(iii), where there are four excited sites: two -2, one -1, and one zero charge. Each of these states are found when sites of the same colour domain are directly adjacent -- in this case, grey to grey. Similarly, Figure \ref{fig:4parter}(b)(iv), shows a +1 charge site where a chain of black meets a chain of grey.

Given the clear instances of excited states and monopole formation in this system -- both in the proposed ground state and where two ground states collide -- it would be of interest to observe their dynamics as the system cools. Similarly, as to whether one domain could dominate via the application of an external magnetic field. In turn, this may allow for tunable control over the location and frequency of monopoles.

\section{Conclusion and future work}

We have presented two systems of rhombic tilings which have been studied as artificial spin ices in a point dipole system. Here, we chose to investigate ASIs decorating an aperiodic, single edge length H$_{\frac{1}{2}\frac{1}{2}}$ tiling ($\psi_{-0.36}$), and compare it to the closely related periodic dice lattice ($\psi_{-1}$). For a view of the systems as a whole, we studied the energy and specific heat as our systems cooled. Close analysis of the final spins on $\psi_{-0.36}$ focused particularly on the energetic states found at the tiling vertices, including the energy levels, excitations, and finite magnetic charges.  Using this analysis, we have also proposed a provisional ground state structure for $\psi_{-0.36}$, which includes excitations at specific 3i and 4i-b sites. Similarly, we explored colliding magnetic domain behaviour, which created additional excited states.

This initial exploration of the  $\psi_{-0.36}$ system provides scope for a number of avenues for future work. Specifically, the nature of the $\psi_{-0.36}$ system's monopoles indicates potential for a tuning of their distribution and frequency, following additional analysis of their dynamics -- the application of an external magnetic field may provide further insight along these lines. Similarly, there is potential for the verification of the ground state using tiling substitution rules and perpendicular space analysis. More generally, both the distribution of vertex types and their local environments strongly influence the system’s energetics and magnetic ordering. Since these factors are governed by the grid-shift parameters defining the $\psi_{-0.36}$ geometry, a clearer understanding of how the shifts control vertex populations and local correlations could allow us to tune these variables deliberately, opening a route to engineering families of systems with novel or advantageous behaviour.

\begin{acknowledgements}
We wish to thank Dr. Liam O'Brien for helpful discussions. SC acknowledges that this work was supported by EPSRC grant EP/X011984/1.
\end{acknowledgements}




\bibliography{ref} 
\clearpage
\appendix
\renewcommand{\thefigure}{A\arabic{figure}}
\setcounter{figure}{0}
\section{Monte Carlo acceptance rate \label{app:acceptance rates}}
\begin{figure}[h] %
	\begin{center}
		\includegraphics[width=0.95\textwidth]{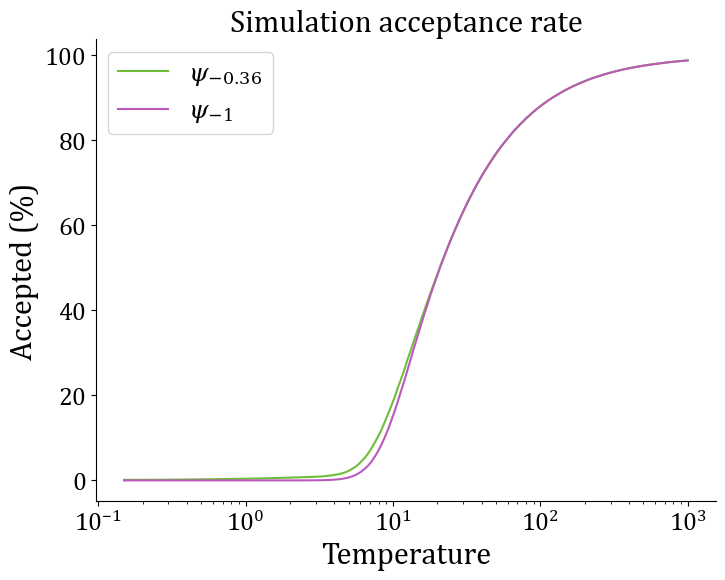} 
	\end{center}
	\caption{Monte Carlo simulation acceptance rates for the $\psi_{-1}$ and $\psi_{-0.36}$ geometry. This is the average total number of spin flips accepted with respect to the total number attempted, calculated for each temperature.}\label{fig:acceptrate}
\end{figure}
\clearpage

\section{Polarised finite charge map of $\psi_{-0.36}$ \label{app:charge}}
\begin{figure}[h] %
	\begin{center}
		\includegraphics[width=0.85\textwidth]{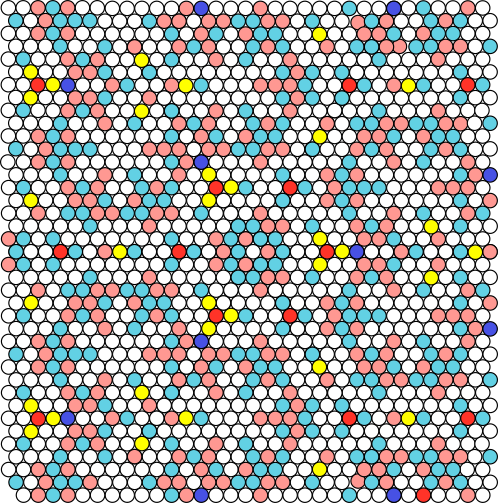} 
	\end{center}
	\caption{The finite charges on each site in the proposed ideal ground state. White represents even vertices with a finite charge of zero; yellow shows excited vertices with a charge of zero; light red and light blue indicates a charge of $\pm$1; and the darkest coloured red and blue indicate $\pm$3 monopoles respectively.}\label{fig:chargemap}
\end{figure}
\clearpage

\section{Colour domain of the $\psi_{-0.36}$ ground state \label{app:groundcolour}}
\begin{figure}[h]
	\centering
	\includegraphics[width = .95\linewidth]{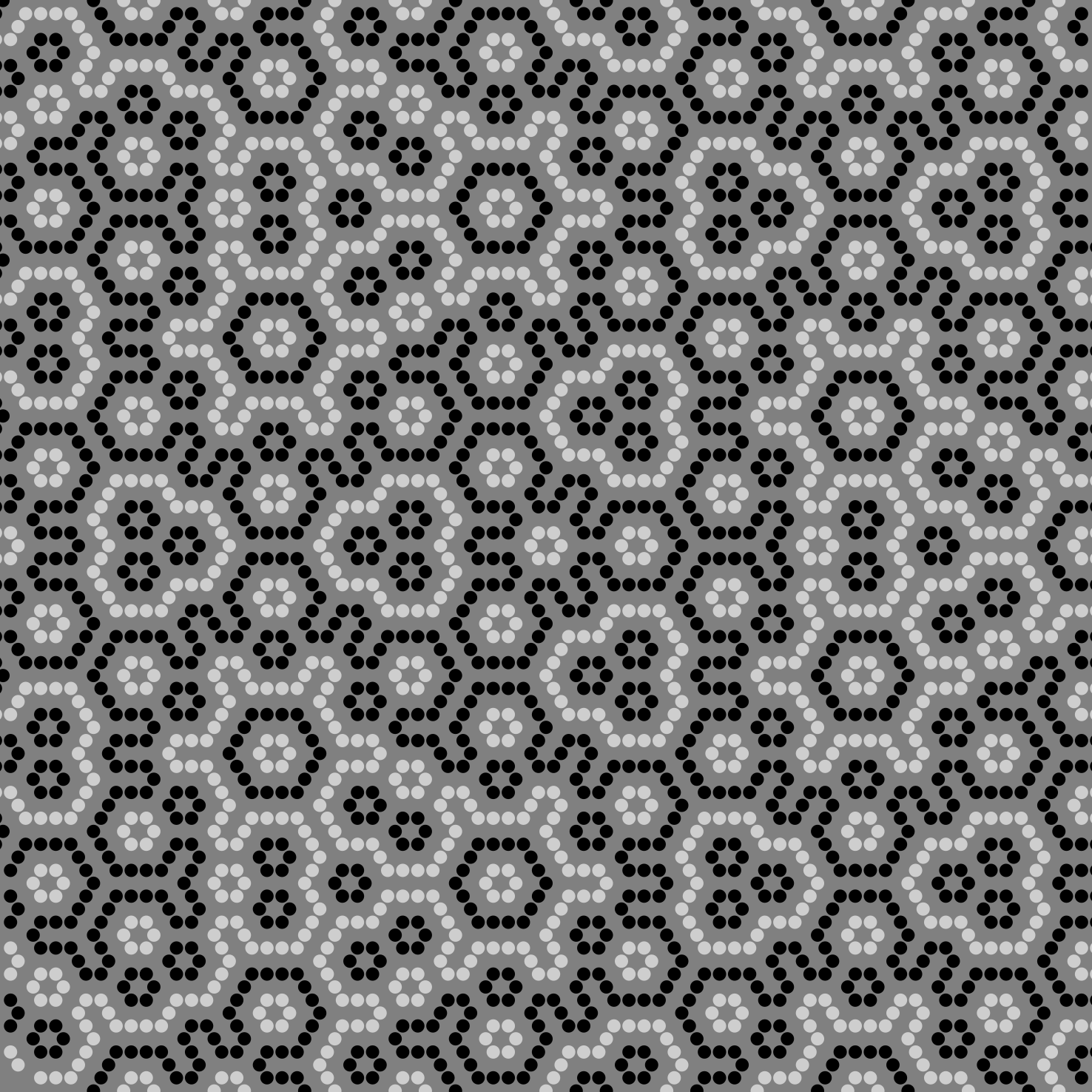}
	\caption{One of the two ground state solutions for $\psi_{-0.36}$, where each circle decorate the mid-edge of the H$_{\frac{1}{2}\frac{1}{2}}$ tiling. Circles are coloured according to the spin direction of the island decorating the tile, as in Figure \ref{fig:4parter}(d).}
\end{figure}

\end{document}